\documentclass[letterpaper, 10pt, conference]{ieeeconf}
\IEEEoverridecommandlockouts
   \usepackage[pdftex]{graphicx}
 
  \graphicspath{{../pdf/}{../jpeg/}}
  \DeclareGraphicsExtensions{.pdf,.jpeg,.png}
  
  \usepackage{amssymb} 
  \usepackage{graphicx}
\usepackage[natural]{xcolor}

\usepackage{changes}

\usepackage{siunitx}

\usepackage[cmex10]{amsmath}
\usepackage[tight,footnotesize]{subfigure}

\usepackage{url}

\usepackage[mathscr]{euscript}
\usepackage{multirow}
\usepackage{comment}

\begin{document}

\title{\LARGE \bf 
Cooperative Tracking of Cyclists Based on\\ Smart Devices and Infrastructure  }

\author{G\"{u}nther Reitberger, Maarten Bieshaar, Stefan Zernetsch, Konrad Doll, Bernhard Sick, and Erich Fuchs
	\thanks{G. Reitberger and E. Fuchs are with the FORWISS, 
			University of Passau, Passau, Germany
			{\tt\footnotesize reitberg@forwiss.uni-passau.de, fuchse@forwiss.uni-passau.de}}
	\thanks{M. Bieshaar and B. Sick are with the Intelligent Embedded Systems Lab, University of Kassel,
		Kassel, Germany
		{\tt\footnotesize mbieshaar @uni-kassel.de, bsick@uni-kassel.de}}
	\thanks{S. Zernetsch and K. Doll are with the Faculty of Engineering,
		University of Applied Sciences Aschaffenburg, Aschaffenburg, Germany
		{\tt\footnotesize stefan.zernetsch@h-ab.de, konrad.doll@h-ab.de}}
}

\maketitle

\begin{abstract}
	
In future traffic scenarios, vehicles and other traffic participants will be interconnected and equipped with various types of sensors, allowing for cooperation based on data or information exchange. This article presents an approach to cooperative tracking of cyclists using smart devices and infrastructure-based sensors. A smart device is carried by the cyclists and an intersection is equipped with a wide angle stereo camera system. Two tracking models are presented and compared. The first model is based on the stereo camera system detections only, whereas the second model cooperatively combines the camera based detections with velocity and yaw rate data provided by the smart device. Our aim is to overcome limitations of tracking approaches based on single data sources. We show in numerical evaluations on scenes where cyclists are starting or turning right that the cooperation leads to an improvement in both the ability to keep track of a cyclist and the accuracy of the track particularly when it comes to occlusions in the visual system. We, therefore, contribute to the safety of vulnerable road users in future traffic.


\end{abstract}



%
\IEEEpeerreviewmaketitle



\section{\large Introduction}
\label{sec_introduction}
\subsection{Motivation}

In our work, we envision a future mixed traffic scenario~\cite{BRZ+17} where traffic participants, such as automated driving cars, trucks, and intelligent infrastructure equipped with sensors, electronic maps, and Internet connection, share the road with vulnerable road users (VRUs), such as pedestrians and cyclists, equipped with smart devices. Each of them itself determines and continuously maintains a local model of the surrounding traffic situation. This model does not only contain information by each traffic participant's own sensory perception, but is the result of cooperation with other traffic participants and infrastructure in the local environment, e.g., based on vehicular ad hoc networks. This joint knowledge is exploited in various ways, e.g., to increase the perceptual horizon of individual road users beyond their own sensory capabilities. Although modern vehicles possess many forward looking safety systems based on various sensors, still dangerous situations for VRUs can occur as a result of occlusions or sensor malfunctions.
Cooperation between the different road users 
can resolve occlusion situations and improve the overall performance regarding measurement accuracy, e.g., 
precise positioning. 

In this article we propose a cooperative approach to track cyclists at an urban intersection robustly and accurately. The cooperatively obtained positional information can then subsequently be used for intention detection~\cite{BZD+17}. In contrast to bare data fusion, cooperation also captures the interactions between different participants. Therefore, we use cooperation as an umbrella term including fusion as an integral part.

\subsection{Main Contributions and Outline}

The main contribution of this article is an approach to cooperatively detect and track 
the position of cyclists at an urban intersection. 
The proposed method incorporates positional information originating from the camera tracks of the 
cyclist's head trajectory as well as velocity and yaw rate estimates originating from a smart device carried 
by the cyclist. This information is adaptively combined using an extended Kalman filtering approach. 
The resulting cooperative tracking mechanism is accurate and, furthermore, it can cope with short term occlusion. The novel metric MOTAP is introduced to evaluate the benefit of cooperation in comparison to a single entity approach.

The remainder of this article is structured as follows: In Sec.~\ref{sec:related_work}, the related work 
in the field of cooperative transportation and tracking methods including smart devices is reviewed.
Sec.~\ref{sec_method_overview} describes the overall approach to cooperatively track cyclists. The methods  and metrics used for evaluation are described in Sec.~\ref{sec_data_acq_eval}. In Sec.~\ref{sec_ResultsOutline}, the experimental results are presented. Finally, in Sec.~\ref{sec_conclusion} the main conclusions and the open challenges for future work are discussed.

\section{Related Work}
\label{sec:related_work}
Many dangerous situations involving vehicles and VRUs occur in urban areas.
The German project Ko-TAG~\cite{Ko-TAG} of the Ko-FAS research initiative \cite{Ko-FAS} aimed to increase the road safety by combining infrastructure-based perception enriched with data from vehicles enabling cooperative perception. Nevertheless, they focused on pedestrians and did not include smart devices.

In~\cite{Thielen2012}, Thielen~et~al. presented a prototype system incorporating a vehicle with the ability of Car-to-X communication and a cyclist with a WiFi enabled smartphone. The authors were able to successfully test a prototype application that warns a vehicle driver if the collision with a crossing cyclist is likely to occur within the next $5\,$seconds. A similar prototype system including Car-to-Pedestrian communication was proposed by Engel et. al. in~\cite{Engel2013}. However, the tracking of the VRU is limited by its positional accuracy due to the usage of smartphone sensors only. It does not make use of a cooperative tracking mechanism. Another approach, combining a radar equipped infrastructure and smart devices in a cooperative way is described by Ru{\ss} et. al. in~\cite{Russ2016}. The radar information is used to correct the global navigation satellite system (GNSS) position data of the smartphone using a simple combination mechanism with fixed weights. Besides a prototype system, the authors did not provide a quantitative evaluation.
In~\cite{MSN17}, Merdrignac et. al. propose a cooperative VRU protection system in which vehicles and pedestrians exchange messages about their position successfully resolving occlusion, i.e., non-line of sight situations. Their proposed system is limited in the real world application due to the necessity of precise smart device localization capabilities, which cannot be provided by the built-in GPS.

In~\cite{BRZ+17}, we presented a cooperative, holistic concept to detect intentions of VRUs by means of collective intelligence, including smart devices carried by the VRU itself. We proposed 
an approach to cooperatively detect cyclists' starting motion and to forecast their future trajectory in~\cite{BZD+17}.
The approach was limited in its application due to the requirement of precise positional information for the trajectory forecast. Particularly, it could not cope with occlusion situations. The cooperative tracking approach presented in this article alleviates this by including smart device information. It can provide a precise VRU position even in the short absence of any visual information.






\section{\large Method}
\label{sec_method_overview}

\begin{figure}[t]
	\centering
	\includegraphics[width=0.48\textwidth, clip, trim=3 90 210 0]{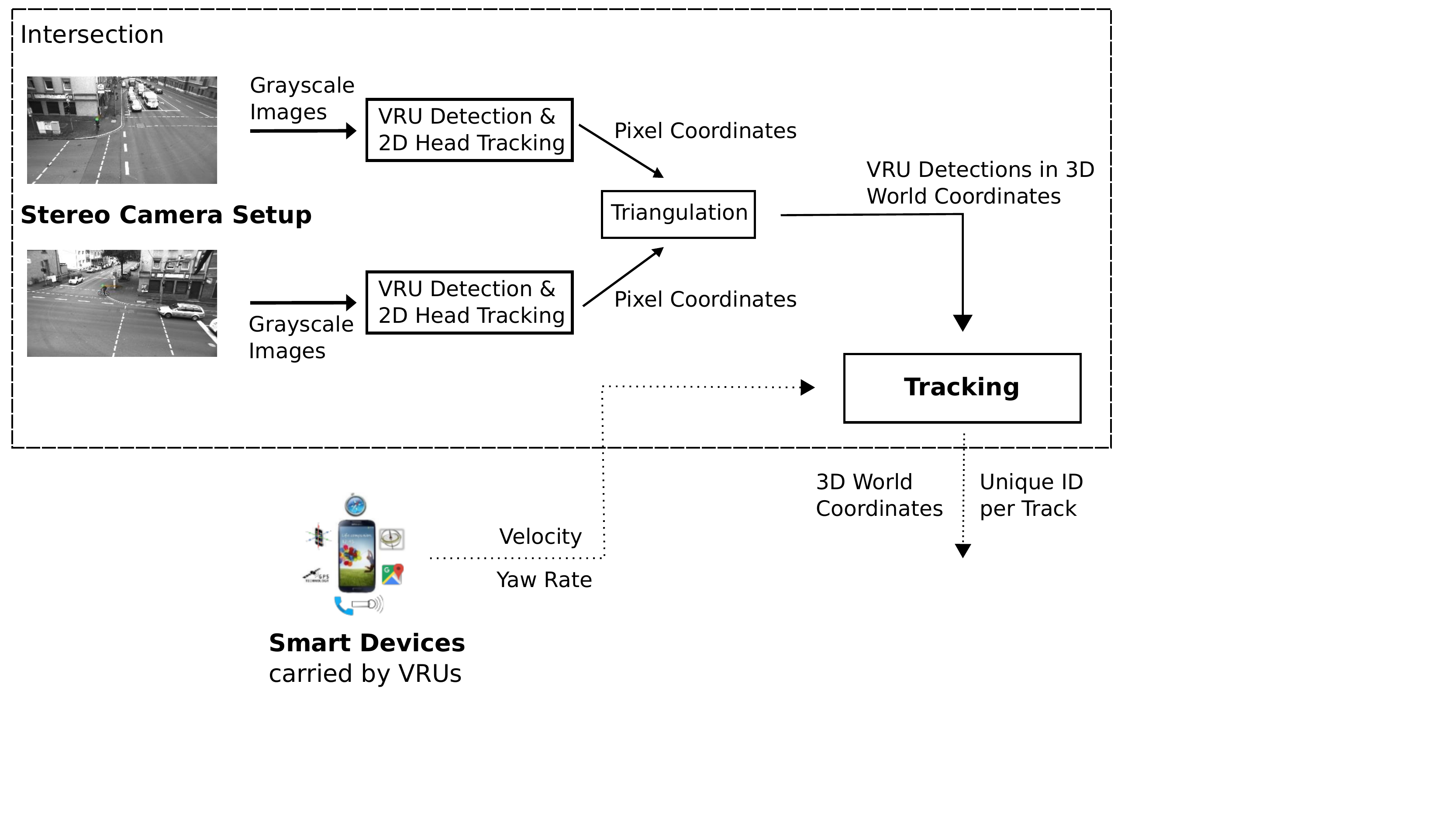}
	\vskip 2mm
	\caption{VRU tracking based on infrastructure and smart devices.}
	\label{fig:CooperativeDetTrack}
	\vskip -0.3cm
\end{figure}

We envision to make use of data provided by all road users including infrastructure in the local environment, allowing to detect VRUs, classify, localize, and track them. Here, we restrict ourself to a research intersection~\cite{GoldhammerIntersection.2012} and smart devices carried by the cyclist.
A schematic of our approach, which illustrates the components and their interaction, is depicted in Fig.~\ref{fig:CooperativeDetTrack}. In the first stage, the cyclist and especially his head is detected in the camera images.
On top of that, a 2D head tracking algorithm is presented to overcome minor detection misses and occlusions. Subsequently, the 3D head position is triangulated using the 2D head position of both camera images. Human activity recognition and machine learning techniques~\cite{Bulling2014THA} based on the smart device inertial measurement unit (IMU) are used to estimate the cyclists yaw rate and velocity. These estimates are sent to the infrastructure, e.g., using an ad hoc network. 
The triangulated head position and velocity and yaw rate estimates are then combined using an extended Kalman filter implementing the cooperative tracking. We focus on tracking the head for two reasons: First, the head is a good indicator for human intentions~\cite{HZD+17}, second, it is in plain view from different camera perspectives and, therefore, perfectly suited for triangulation. Moreover, the integration of smart device based velocity and yaw rate estimates allows to track a cyclist even in the absence of any visual information. Strong head movements can lead to differences in the velocity induced by the head detections and measurements of body worn smart devices. We do not treat this issue explicitly in this work, but we are able to handle it in our setting, because strong head movements are visible in the energy of the smart devices and we attach smart devices at the helmets to detect head motions. 


For the communication between the smart devices and the infrastructure, we assume that it is realized by means of an ad hoc network. The approach assumes an idealized communication medium without any considerable communication delays and synchronized devices using GPS timestamps. 




\subsection{Image based Cyclist Detection}
\label{subsec:method:vru_detection}

A setup of two high definition cameras mounted in a wide stereo angle at opposite corners of the intersection forms one part of the cooperating agents. We perform image based cyclist detection on every camera. 


The detector development is performed with the state of the art TensorBox framework described in \cite{Tensorbox}. The framework enables, in a comfortable way, training of neural networks to detect objects in images using a classifier of ones choice embedded in the architecture described in \cite{Russ2015ObDet}. As a classifier we use the default GoogLeNet~\cite{GoogLeNet}. The proposed architecture is, although a generic one, especially applicable to person detection in crowded scenes as it directly generates a set of object bounding boxes as an output and aims to make the post processing in form of merging and non-maximum suppression to avoid multiple detections obsolete.

\begin{figure}[ht]
	\centering
	\includegraphics[width=0.48\textwidth, clip, trim=0 270 320 0]{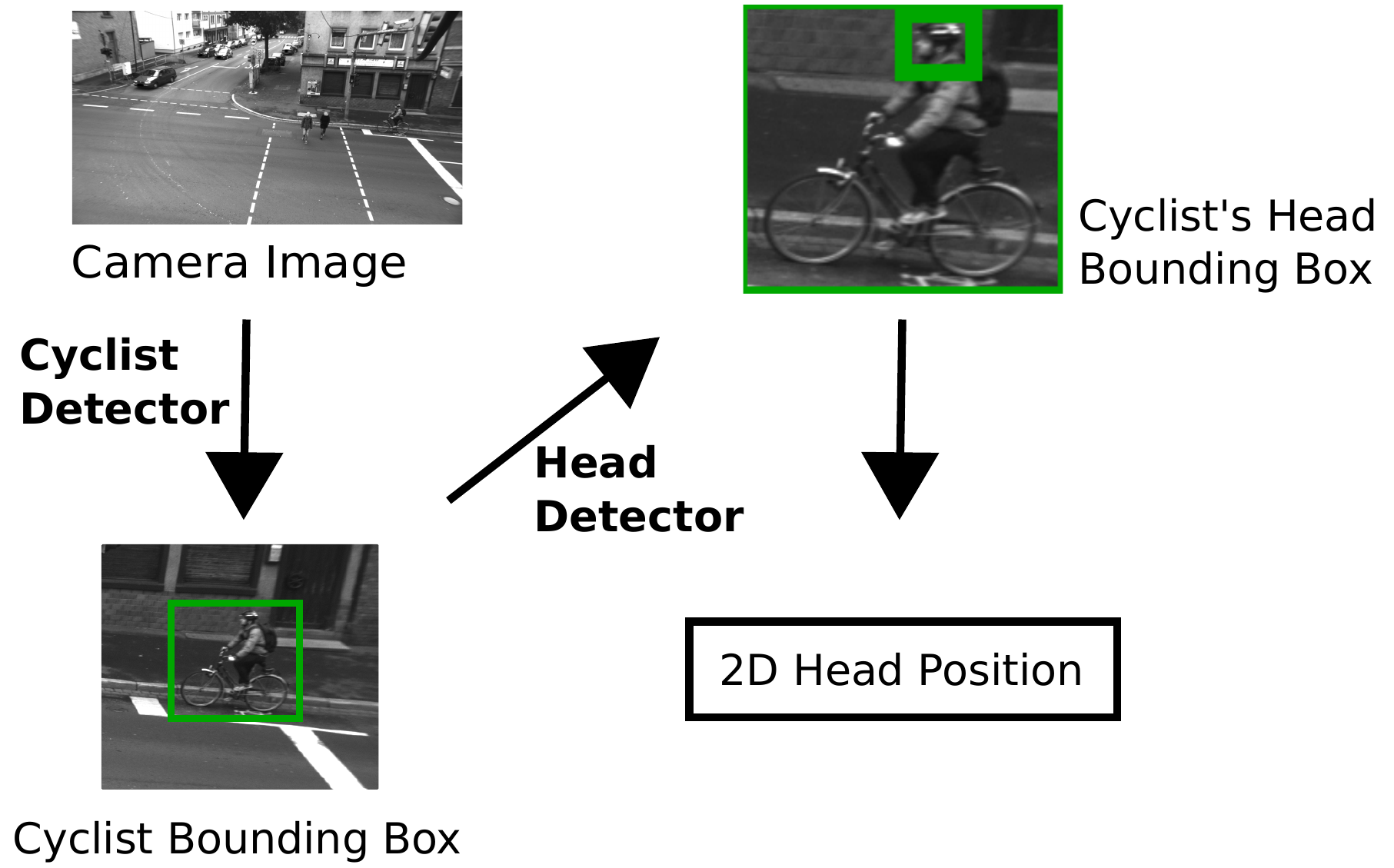}
	\vskip 2mm
	\caption{Cyclist's head detection using self trained cyclist and head detectors.}
	\label{fig:HeadDetection}
	\vskip -0.3cm
\end{figure}

As we are interested in tracking cyclists via the center of their heads, we trained two detectors. Fig.~\ref{fig:HeadDetection} illustrates the detection process. At first, a cyclist detector is essential. To generate bounding boxes around the cyclists as training data, the bike detector proposed by Felzenszwalb et al. in~\cite{Felzenszwalb2010} is applied to a sufficiently big region of interest around the labeled head position. The head detector was trained in an analogue way. This time, only the calculated bounding box surrounding the labeled cyclist was used as input image. The trained detector performs on cyclist bounding boxes and produces bounding boxes for the heads of the cyclists. The output of the detection algorithm is a head position that is a simple determination of the center of the bounding box produced by the head detector. 

Due to changing weather and illumination situations or simply short (partly) occlusions detection misses are unavoidable. To reduce the number of such detection misses, a constant velocity (CV) Kalman filter (KF)~\cite{Bar-Shalom2001} in combination with a memory functionality is implemented. The KF operates on the state space $[u, v, \dot{u}, \dot{v}]$, with $u$ and $v$ being pixel coordinates and $\dot{u}$ and $\dot{v}$ being the corresponding derivatives in time. To solve the detection to track assignment, the Munkres algorithm~\cite{Munkres1957} is used. If there is a detection with no track assigned, because it is more than $40$ pixels in Euclidean distance away from every existing track, a new KF track is started. If there is a track with no detection assigned, an internal detection miss counter is increased. If the ratio of the miss counter to the total age of the track exceeds $30\%$, the track is considered lost and gets deleted. A track is also considered lost, when there has not been an update for one second. To make the system more robust, a track has to have at least an age of four frames to be considered as valid. This introduces some delay, but reduces the number of false positives. The output of the combined 2D detection and tracking is a number of tracks. The current position in pixel coordinates of each track is interpreted as detection and considered in the following triangulation.


The wide angle setup of the cameras at the intersection allows for determination of 3D coordinates via triangulation. It is designed for a spatial resolution better than \SI{10}{\centi\meter}~\cite{GoldhammerIntersection.2012}. The 3D coordinates are important in our cooperative setting to exchange absolute information. The calculation of triangulation follows the basic knowledge of epipolar geometry as it can be found in~\cite{Hartley2003}. 

\subsection{Yaw Rate and Velocity Estimation using Smart Devices}
\label{subsec:method:smart_device}

\begin{figure}[h]
	\centering
	\includegraphics[width=0.48\textwidth, clip, trim=16 15 100 17]{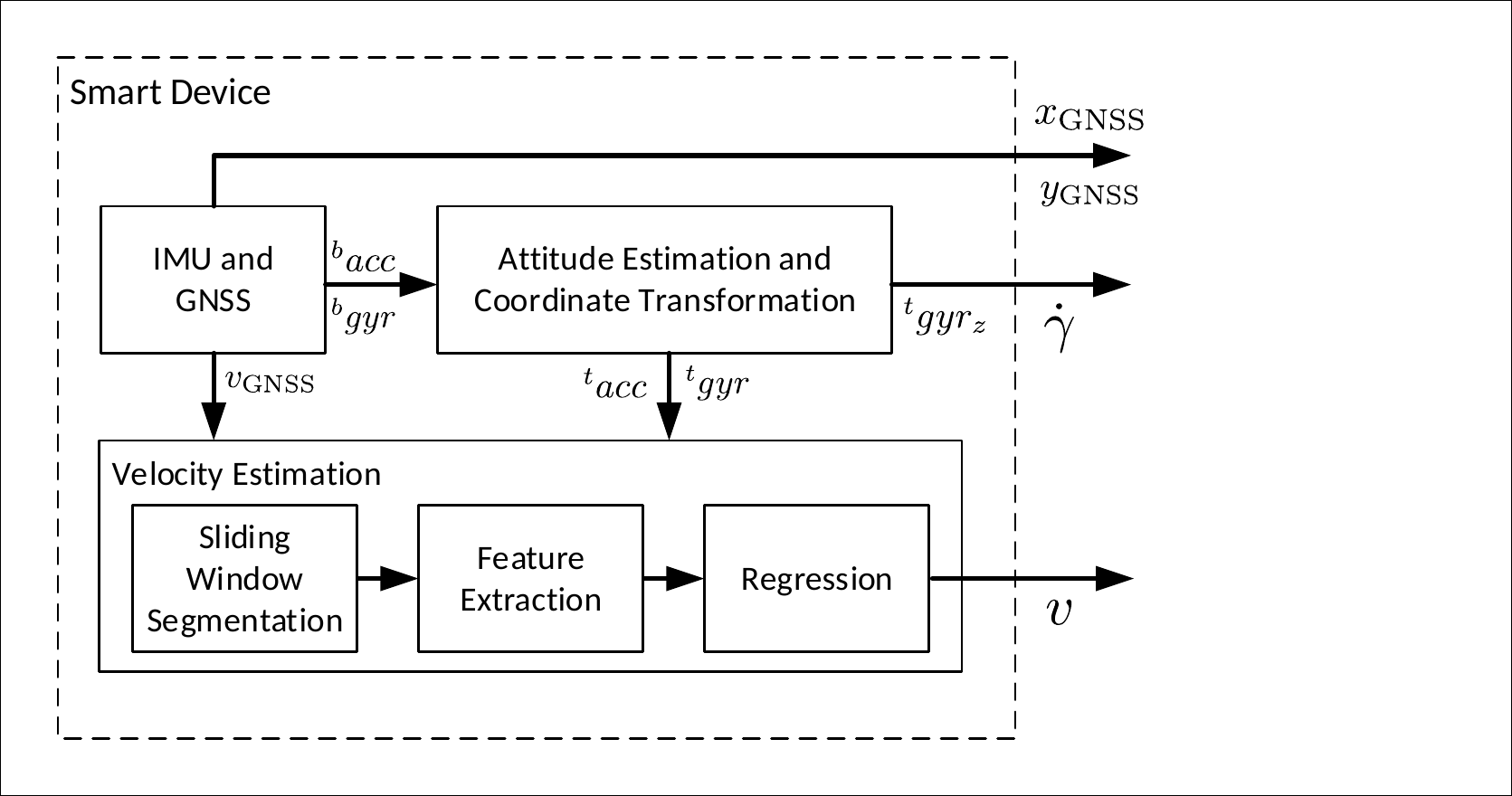}
	\vskip 2mm
	\caption{Process of smart device based yaw rate and velocity estimation. The upper blocks include the GNSS and IMU based attitude estimation, used for transformation of the measurements into the local tangential frame. The lower block depicts the human activity pipeline used for velocity estimation.}
	\label{fig:velocity_estimation}
	\vskip -0.3cm
\end{figure}

In this section the yaw rate and velocity estimation using smart devices is described.
Besides inertial measurements, i.e., the accelerometer and gyroscope sensor, also position and velocity information by the GNSS is nowadays available on nearly every smart device.
Inertial navigation systems (INS)~\cite{TW04} are widely used in aerospace and automotive industry, e.g., for dead reckoning.
Here, first the attitude is estimated and then subsequently the velocity and position are obtained by integration. These algorithms are not directly applicable for smart devices carried by pedestrians and cyclists as small errors in the attitude calculation, due to relative high ego motion, e.g., cyclists pedaling, and low-cost inertial sensors, accumulate, deteriorating the velocity or position estimation. 
In order to be more robust against errors in the attitude estimation, our approach for velocity estimation is realized by means of human activity recognition techniques~\cite{Bulling2014THA} complemented by velocity measurements originating from the GNSS integrated in the device. A schematic of the approach is depicted in Fig.~\ref{fig:velocity_estimation}.

We consider the yaw rate $\dot{\gamma}$ measurements and the velocity estimation $v$ in the local tangential frame $t$, i.e., an arbitrary local coordinate frame whose $z$-axis points toward the sky and is perpendicular to the local ground plane. The velocity $v$ is defined as the magnitude of the velocity $v_x$ and $v_y$ in the local tangential frame.
We assume that the cyclist is always moving in forward direction and ego-motion resulting in 
an increased velocity magnitude, e.g., small side steps are negligible. By considering only the magnitude of the velocity and the yaw rate (i.e., angular velocity around the $z$-axis), there is no need to estimate the transformation of the device with respect to a global coordinate frame. Moreover, we do not need a compass which is sensitive to a precise calibration~\cite{MGF+17}.

The acceleration ${}^{b}acc$ and gyroscope ${}^{b}gyr$ measurements are obtained in the body coordinate frame $b$. The transformation between $b$ and the local tangential plane $t$, i.e. ${}^{b}acc$ and ${}^{b}gyr$ to ${}^{t}acc$ and ${}^{t}gyr$, is obtained by estimating the local gravity vector, which is supplied by nearly all modern mobile systems. Therefore, we assume this transformation as provided. The approach presented here uses features computed from accelerometer and gyroscope sensors sampled with a frequency of \SI{50}{\Hz}. 
The smart device's integrated GNSS (position $x_{\text{GNSS}}$, $y_{\text{GNSS}}$, and velocity $v_{\text{GNSS}}$ in moving direction) is sampled with \SI{1}{\Hz}, i.e., the maximal frequency provided by current smart devices (Android and iPhone).

%

We assume that the cyclist's motion with respect to the rotation around the $z$-axis of the local tangential frame is negligible. Therefore, we can use the rotation ${}^{t}gyr_z$, i.e., rotation around the $z$-axis, as yaw rate $\dot{\gamma}$ estimate. In order to reduce the effect of the ego-motion by the smart device (e.g., induced by leg movement), we low-pass filter the gyroscope ${}^{t}gyr_z$ with window size \SI{0.25}{\second}.


The velocity estimation is realized by a machine learning approach based on ${}^{t}acc$ and ${}^{t}gyr$.
Orientation-independence is achieved by considering the magnitude of the accelerometer and gyroscope values in the local horizontal $x-y$ plane. Moreover, the projection of the sensor values on the local vertical $z$-axis, i.e., the gravity axis, is considered.
A sliding window segmentation of window sizes \SI{1}{\second} is performed on each of the transformed signals and features, such as the mean and energy, are computed. These features are used, since calculating for example the mean of the acceleration is directly related to the velocity. Additionally, the magnitude of the discrete Fourier transform (DFT) coefficients are also considered as input features, as successfully applied for human walking speed estimation in~\cite{PPC+12}. The coefficients are normalized with respect to the overall energy in the respective window. As in~\cite{PPC+12}, the window size is set to \SI{5.12}{\second} and coefficients up the $5^{\mathrm{th}}$ order are considered. 

The features based on activity data capture the dependency between pedaling frequency and the velocity well, 
but can not model the dependency on the engaged gear. 
Therefore, we additionally consider the velocity provided by the 
smart device integrated GNSS. We calculate features based on the coefficients of a third order orthogonal polynomial expansion~\cite{Fuchs2010} using a sliding window size of \SI{5.0}{\second}. The coefficients are 
in a least-squares sense best estimators of the signal's slope and curvature in the approximating window.
These input features are up-sampled to \SI{50}{\Hz} using a zero-order hold filter.

The velocity estimation is realized by means of a frame-based random forest regression~\cite{Breiman2001} at discrete points with a frequency of \SI{50}{\Hz}. 
The regression model is trained with sample velocity data originating from manually labeled and additionally 
smoothed head trajectories. The model is trained with $300$ decision trees with a maximal tree depth of six.
By considering the mean squared deviation of each regression tree from the ensemble average prediction, we 
obtain an estimate of the variance representing the uncertainty of the regression forest $\sigma_{v}^2$. 
GNSS requires the availability of satellite signals, which is especially in urban areas not always given or noisy due to multipath effects. For the case of GNSS outage, we train another random forest regression, but this 
time without GNSS based features. The prediction's variance is slightly increased, especially for fast moving cyclists.

\subsection{Cooperative Tracking}
\label{subsec:method:coop_tracking}

So far, we have presented, how we attain the 3D coordinate positions of cyclists moving in the field of view of the cameras installed at the intersection and how we extract velocity and yaw rate data from the smart devices of the observed cyclists. As the evaluation is focused on the $x$ and $y$ coordinates of the cyclists, the modeling of the $z$ coordinate in form of a constant velocity approach is left out in the following for simplicity reasons. To combine velocity, yaw rate and 2D coordinate positions, we set up an extended Kalman filter (EKF)~\cite{Bar-Shalom2001} with the state space $[x, y, \gamma, \dot{\gamma}, v]^T$ with $x, y$ being the coordinates describing the position of the cyclist, $\gamma$ the yaw, $\dot{\gamma}$ the yaw rate and $v$ the absolute velocity in the direction of movement. The corresponding state transition for a time step $T$ at a state $\mathbf{x}$ is given by
\[ f(\mathbf{x}) := \begin{bmatrix}
x + cos(\gamma)\, a - sin(\gamma)\, b \\
y + sin(\gamma)\, a + cos(\gamma)\, b \\
\gamma + \dot{\gamma}\, T\\
\dot{\gamma}\\
v\\
\end{bmatrix}\]
with $a =\frac{sin(\dot{\gamma}\, T)\, v}{\dot{\gamma}}$ and $b = \frac{(1 - cos(\dot{\gamma}\, T))\, v}{\dot{\gamma}}$. The motion model is the bike model adapted from the work by Bar-Shalom et al. in~\cite{Bar-Shalom2001}. To linearize the non-linear model, the EKF uses the \textit{Jacobian} $F$ of the state transition function $f$. The process noise within a time step $T$ is modeled as a constant acceleration in the direction of movement and a constant offset of the yaw rate. The noise $\mathbf{w} = [w_{\dot{\gamma}}, w_{\dot{v}}]^T$ is assumed to be a zero mean multivariate Gaussian with covariance $Q_{\mathbf{w}} = diag[\sigma_{w_{\dot{\gamma}}}^2, \sigma_{w_{\dot{v}}}^2]$. The state transition for a state $\mathbf{x}$ considering the modeled noise is the following:
\[\mathbf{g}(\mathbf{x}, \mathbf{w}) := \begin{bmatrix}
x + cos(\gamma)\, a - sin(\gamma)\, b \\
y + sin(\gamma)\, a + cos(\gamma)\, b \\
\gamma + (\dot{\gamma} + w_{\dot{\gamma}})\, T\\
\dot{\gamma} + w_{\dot{\gamma}}\\
v + w_{\dot{v}}\, T\\
\end{bmatrix}\quad \text{, with}\]
\[a = \frac{(0.5\, T\, w_{\dot{v}} + v)\, sin(T\, (\dot{\gamma} + w_{\dot{\gamma}}))}{\dot{\gamma} + w_{\dot{\gamma}}}\quad \text{and}
\]
\[b = \frac{(0.5\, T\, w_{\dot{v}} + v)\, (1-cos(T\, (\dot{\gamma} + w_{\dot{\gamma}})))}{\dot{\gamma} + w_{\dot{\gamma}}}\ .
\]
The derivative of $\mathbf{g}(\mathbf{x}, \mathbf{w})$ of the noise $\mathbf{w}$ evaluated at $\mathbf{w} = \mathbf{0}$ results in the linearized Matrix $\Gamma(\mathbf{x})$. This matrix describes the noise gain in the EKF setting. The process noise covariance matrix $Q$ is calculated by $\Gamma(\mathbf{x})\, Q_{\mathbf{w}}\, \Gamma(\mathbf{x})^T$, as it is described in~\cite{Bar-Shalom2001}. In our setting, $T$ is fixed by \SI{20}{\milli\second}. \cite{Bar-Shalom2001} suggests to choose $\sigma_{w_{\dot{v}}}$ between $0.5\, a_{max}$ and $a_{max}$. Therefore, $\sigma_{w_{\dot{v}}}$ is set to \SI{2.5}{\meter\per\second}, a high acceleration/deceleration for a cyclist, and $\sigma_{w_{\dot{\gamma}}}$ to \SI{1.5}{\radian\per\second}, accounting for a big change of yaw rate over one second.

We consider three measurement models. The first one performs an update with both the position and the smart device data, the second one with the smart device data only, and the third one with position only. This covers all possible states of information per time stamp. If there is no information at a time stamp, no update can be done. The standard deviations for the measurement noise are given by \SI{0.15}{\meter} for $\sigma_x$ and $\sigma_y$ and \SI{0.3}{\radian\per\second} for $\sigma_{\dot{\gamma}}$. They were estimated by comparison with the ground truth data. Considering $\sigma_v$, the estimation by the regression model from Sec.~\ref{subsec:method:smart_device} is used. As the measurement errors in $v$ and $\dot{\gamma}$ are estimated per time step, an additional division by $T$ is necessary, leading to the measurement noise covariance matrix $R = diag[\sigma_x^2, \sigma_y^2, (\sigma_{\dot{\gamma}}/T)^2, (\sigma_{v}/T)^2] $. By low pass filtering the gyroscope, estimating the velocity over a sliding window, and calculating the 3D positions on Kalman filtered detections, a form of auto correlation is induced using these as measurements in the above EKF. So far, this issue has not been modeled.

Following the idea of the already presented tracking in the image space, we want to overcome situations of missing data by a memory functionality. The same algorithm as in Sec.~\ref{subsec:method:vru_detection} is also used in the 3D scenario with the following differences in parameters. If a detection is more than \SI{2}{\meter} away from a track, it is not considered for an assignment in the Munkres algorithm anymore. A track is lost, when there has not been an update in position for more than \SI{2}{\second} or the miss ratio exceeds \SI{50}{\percent}. 

Additionally, the assignment of the smart device data to the corresponding track has to be solved. Therefore, the distance of a measurement to an existing track has to be evaluated. Let $\mathbf{z} = [\dot{\gamma}, v]^T$ be a new measurement by the smart device, then $\mathbf{y} := \mathbf{z} - H\, \mathbf{x_p}$ defines the measurement residual of the predicted state $\mathbf{x_p}$ of a track and $\mathbf{z}$. The measurement matrix $H$ simply extracts $\dot{\gamma}$ and $v$ from $\mathbf{x_p}$. The Mahalanobis distance is defined as $\sqrt{y^T\, S^{-1}\, y}$ with $S = H\, P\, H^T + R$ being the innovation covariance matrix that is calculated in the update step of the EKF by the predicted covariance matrix $P$, $H$, and the measurement noise covariance matrix $R$ as defined above. The Mahalanobis distance measures the length of the residual in standard deviations. The disadvantage is that large predicted covariances can lead to small distances and measurements are more likely assigned to tracks with large uncertainties. To cope with high uncertainties, a penalty term gets added and the final penalized Mahalanobis distance measure is the following:
\[d(H\, \mathbf{x_p}, \mathbf{z}, S) = \sqrt{y^T\, S^{-1}\, y + \ln(det(S))}\]
A more detailed derivation can be found in~\cite{Altendorfer2016}. In our case, we only have a single smart device source that has to be assigned to potentially multiple tracks. The assignment is solved by a nearest neighbor approach based on the penalized Mahalanobis distance.


\section{Data Acquisition and Evaluation}
\label{sec_data_acq_eval}

\subsection{Data Acquisition}
\label{subsec_data_acq}
The developed tracking algorithm is evaluated in experiments conducted with 52 female and male test subjects in the age between $18$~-~$54$. The test subjects were equipped with a Samsung Galaxy S6 smart device carried in the trouser front pocket and instructed to move between certain points at an intersection while following the traffic rules. The recorded scenes included waiting, starting, driving through, and turning (left, right) behavior. To record the cyclist trajectories, a wide angle stereo camera system consisting of two high definition cameras ($1920 \times 1080\,$px, $50\,$fps)~\cite{GoldhammerIntersection.2012} was used. The timestamps of the smartphone and the research intersection are synchronized offline. The head tracks on the video cameras are labeled by human operators and assumed to be close to the ground truth. The labeled positions are triangulated to obtain 3D coordinates.

\begin{figure} [h]
	\centering
	\includegraphics[width=0.40\textwidth, clip, trim=120 270 200 70]{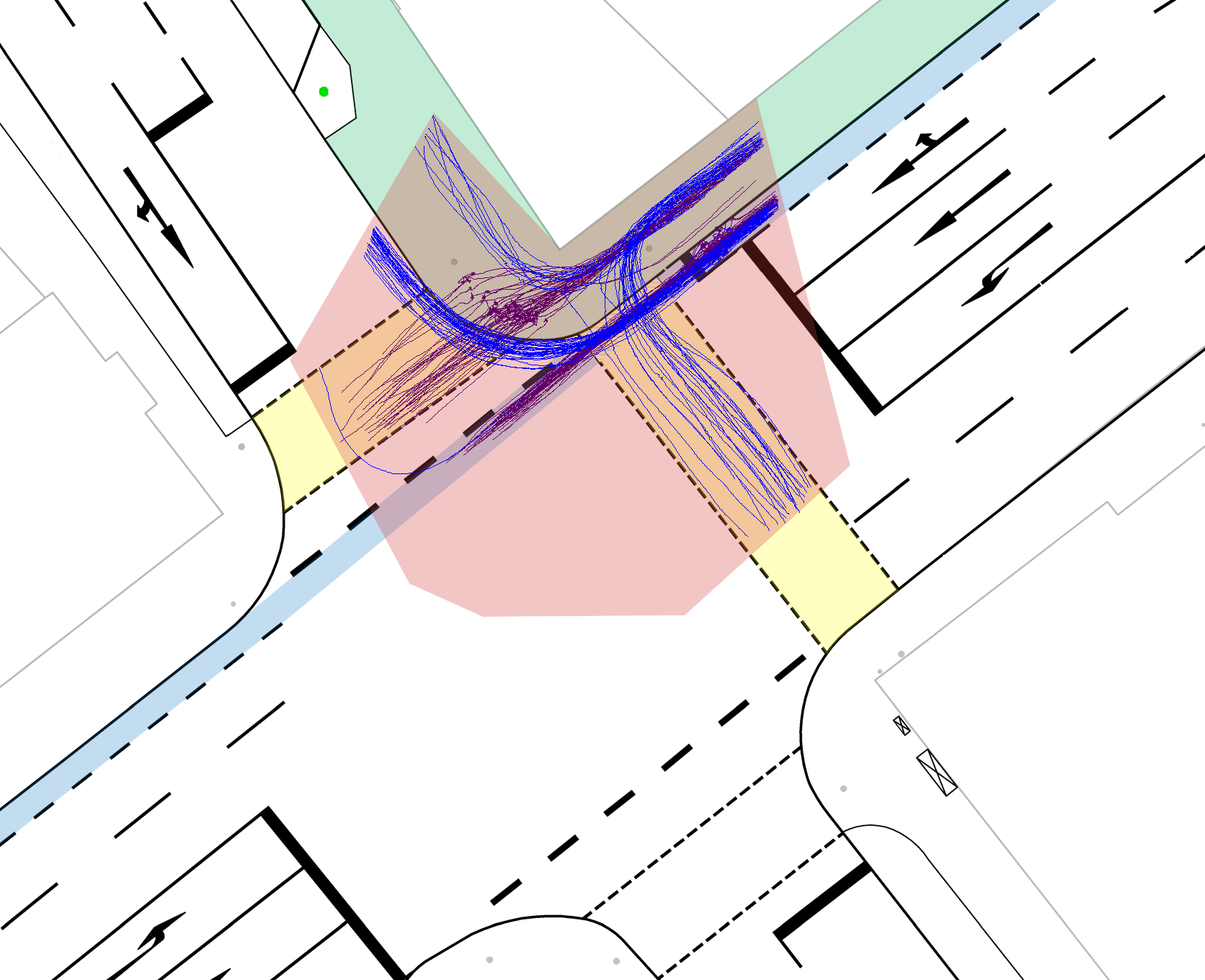}
	\vskip 3mm
	\caption{Overview of the intersection with all cyclists' trajectories.
		The turning right tracks are blue, whereas the starting ones are purple.}
	\label{fig:trajectories}
	\vskip -0.3cm
\end{figure}

\subsection{Evaluation}
\label{subsec_evaluation} 

In total 74 turning right and 87 starting scenes are fully labeled, processed, synchronized and thus available for evaluation. The extracted trajectories are plotted in Fig.~\ref{fig:trajectories}. Blue ones represent the turning right scenarios, whereas purple ones visualize the starting scenes. The intersecting field of view of both cameras is sketched in light red. Starting scenes are designed in such a way that the test subjects approach red traffic lights. They have to stop and start in a straight direction, when the lights turn green again. This should ensure a natural starting behavior. For the evaluation, only the process after the stopping at the red lights is considered. In the case of turning right, the test subject may as well stop at red lights before turning right or be in motion throughout the complete scene. To cut off the waiting, only the last \SI{12}{\second} of a scene were used.

The use case of our cooperative approach are scenes, where occlusions compromise a proper tracking of cyclists. 
If the 2D position information by one camera is missing, there is no triangulation possible anymore. Therefore, there is no 3D position, as well. We create artificial occlusions of \SI{1}{\second} and \SI{2}{\second} duration by dropping detections in one camera. Occlusions in accelerating or direction changing motions are the most interesting, because they hide crucial information for tracking. We thus aim to place the artificial occlusions in such states. The recorded scenes end shortly after performing starting or turning, as the cyclists leave the camera view without stopping. Therefore, the occlusions were defined in a fixed temporal distance to the last frame. The \SI{1}{\second} occlusion starts at the same frame like the \SI{2}{\second} one.

We will compare the trajectories created by the intersection only model with the ones by the smart device integrating model. In the field of object tracking the multiple object tracking precision (MOTP) and multiple object tracking accuracy (MOTA) metrics are established. In~\cite{Bernardin2008}, they are defined for the multi-object tracking scenario. In our setting we only have one ground truth trajectory per scene. Therefore, we define MOTP (in an adapted version) and MOTA for the single object tracking task
\begin{equation}
	\label{equ:MOTP}
	\text{MOTP} := \frac{(\sum_t d_t) + (\sum_t lm_t)*\tau}{(\sum_t c_t) + (\sum_t lm_t)}
\end{equation}
\begin{equation}
	\label{equ:MOTA}
	\text{MOTA} := 1 - \frac{\sum_t (dm_t + 2*lm_t)}{\sum_t g_t}
\end{equation}
with $\delta_t$ being the Euclidean distance of the modeled track to the ground truth track at time $t$, $\tau$ being the maximum distance, a track gets assigned to the ground truth,
\[d_t = \begin{cases}
\delta_t, & \text{if } \delta_t <= \tau\\
0, & \text{otherwise}
\end{cases}\quad \text{, and}\]
\[c_t = \begin{cases}
1, & \text{if } \delta_t <= \tau\\
0, & \text{otherwise}
\end{cases}\quad .\]
If $c_t$ equals $0$, it is called a miss, as the track misses to model the object. The variable $g_t$ is $1$, if a ground truth label exists at time $t$ and $0$ otherwise. The variable $dm_t$ is $1$ at time $t$, if there is no track at all, i.e., a detection miss, whereas $lm_t$ is $1$ and counts a localization miss, if a track exists, but the distance $\delta_t$ is bigger than $\tau$.

MOTP is used to measure how accurately a track follows the ground truth, if a track exists. If the track distance to the ground truth exceeds the threshold $\tau$, it gets penalized by $\tau$. MOTA penalizes missing tracks alone, not accounting for any distances. Both have to be considered to assess the quality of a track. At the same time, minor differences in MOTP or MOTA do not indicate a significantly better or worse track. Therefore, the significance thresholds $\alpha$ for MOTA and $\beta$ for MOTP are introduced and track $A$ is considered \textit{better performing} than track $B$, if the condition
\begin{equation}
(\text{MOTA}_A > \text{MOTA}_B + \alpha) \wedge (\text{MOTP}_A < \text{MOTP}_B + \beta)
\label{equ:MOTAP_1}
\end{equation}
or the condition
\begin{equation}
(\text{MOTA}_A > \text{MOTA}_B - \alpha) \wedge (\text{MOTP}_A < \text{MOTP}_B - \beta)
\label{equ:MOTAP_2}
\end{equation}
holds. We define the new metric
\[\text{MOTAP}_{\alpha, \beta}(A, B) := 
\begin{cases}
1, & \text{if condition } \ref{equ:MOTAP_1} \text{ or } \ref{equ:MOTAP_2} \text{ holds }\\
0, & \text{otherwise}
\end{cases}\]
combining MOTA and MOTP to have a single measure to rank the quality of two tracking algorithms regarding one test scene.

\section{Experimental results}
\label{sec_ResultsOutline}

In this section, we compare the position only tracking model, referred $\mathscr{P}$, with the one combining positional and smart device data, referred $\mathscr{C}$. We evaluate in several test runs on both starting and turning right scenes the ability of the specific tracking models to follow the ground truth track. As the smart device data only affects the $x$ and $y$ coordinates of the tracks and we want to investigate the effect of adding smart device information, the distances to the ground truth track are only evaluated regarding the $x$ and $y$ coordinates.

Tab.~\ref{tab:noOcclusionEval} presents MOTP, given in meters, and MOTA for the miss threshold $\tau = \SI{1}{\meter}$ and no artificial occlusion using the characteristic numbers minimum, maximum and mean. MOTAP is calculated with $\alpha = 0.025$ and $\beta = 0.01$.

The choice of $\tau = \SI{1}{m}$, meaning a miss is counted, if the distance of a track to the ground truth exceeds \SI{1}{\meter}, is quite a standard choice in object tracking~\cite{Milan2013}. The value for $\alpha$ is intended to be a small threshold and $\beta$ is intentionally quite high relative to the mean performance to lay more weight on MOTA as it determines, if the object is tracked at all.

{
	\setlength{\tabcolsep}{3pt}
	\setlength{\abovecaptionskip}{5pt plus 3pt minus 3pt}
	\begin{table}[htb]
		\begin{center}
			\caption{Evaluation of all scenes without occlusions.} 
			\begin{tabular}{|*{8}{c|}}
				\hline
				Scenes & \multicolumn{3}{c|}{$\text{MOTP}_\mathscr{P}$} & \multicolumn{3}{c|}{$\text{MOTA}_\mathscr{P}$} & $\text{MOTAP}(\mathscr{P}, \mathscr{C})$ \\\hline
				& $max$ & $min$ & $mean$ & $max$ & $min$ & $mean$ & $\Sigma$ \\\hline
				Starting & 0.311 & 0.029 & 0.071 & 1 & 0.206 & 0.974 & 0 \\\hline
				Turning & 0.326 & 0.038 & 0.084 & 1 & 0.296 & 0.914 & 0 \\\hline
				 & \multicolumn{3}{c|}{$\text{MOTP}_\mathscr{C}$} & \multicolumn{3}{c|}{$\text{MOTA}_\mathscr{C}$} & $\text{MOTAP}(\mathscr{C}, \mathscr{P})$ \\\hline
				Starting & 0.215 & 0.029 & 0.065 & 1 & 0.206 & 0.980 & 7 \\\hline
				Turning & 0.306 & 0.038 & 0.080 & 1 & 0.344 & 0.922 & 5 \\\hline								
			\end{tabular}
			\label{tab:noOcclusionEval}
		\end{center}	
		\vspace{-3mm}	
	\end{table}
}

One can see, that for both data sets, the models perform with an average precision of below \SI{10}{\centi\meter} and an average MOTA score above \SI{90}{\percent} without any artificial occlusions. The turning scenes are more challenging, as both MOTA and MOTP scores are worse in average. The two models operate on a comparable performance regarding the mean values of MOTP and MOTA with slight advantages for model $\mathscr{C}$, as both mean values are better for the two scene types. This can also be seen in the scene wise comparison via MOTAP, as model $\mathscr{C}$ slightly outperforms model $\mathscr{P}$. Regarding the starting scenes, there is no scenes in which $\mathscr{P}$ performs better than $\mathscr{C}$, but $7$ vice versa. Considering the turning scenes, it is zero against five.

{
	\setlength{\abovecaptionskip}{5pt plus 3pt minus 3pt}
	\begin{table}[htb]
		\begin{center}
			\caption{MOTAP of scenes under artificial occlusions.} 
			\begin{tabular}{| c | c | c | c |}
				\hline
				Scene Type & Occlusion[s] & $\Sigma\, \text{MOTAP}(\mathscr{P}, \mathscr{C})$ & $\Sigma\, \text{MOTAP}(\mathscr{C}, \mathscr{P})$ \\ \hline
				Starting & 1 & 8 & 18 \\\hline
				Starting & 2 & 19 & 30 \\\hline
				Turning & 1 & 3 & 33 \\\hline
				Turning & 2 & 9 & 49 \\\hline
			\end{tabular}
			\label{tab:MOTAP_OcclusionEval}
		\end{center}	
		\vspace{-3mm}
	\end{table}
}

The scenes evaluated in Tab.~\ref{tab:MOTAP_OcclusionEval} contain the artificial occlusions defined in Sec.~\ref{subsec_evaluation} in addition to the natural detection misses. Focusing on the starting scenes, the combined model performs better in $18$ scenes for the \SI{1}{\second} occlusion case and in $30$ for the \SI{2}{\second} ones. Model $\mathscr{P}$ performs better in $8$ respectively $19$ scenes. With the turning scenes, the difference is bigger. Model $\mathscr{C}$ outperforms model $\mathscr{P}$ with $33$ over $3$ and $49$ over $9$. The velocity estimation on the smart devices is more imprecise during acceleration and deceleration. Moreover, when there is no change in direction, the position only model has no disadvantage under occlusion regarding the yaw rate. It may even have an advantage, when the yaw rate is corrupted. Therefore, the advantage of the combined model is reduced in the starting scenes. In the turning scenes, nevertheless, the combined model leads to a real improvement in over half of the scenes for \SI{2}{\second} of occlusion. There are still scenes, in which model $\mathscr{P}$ performs better. This is due to corrupted and imprecise smart device data.


\begin{figure}[htb]
	\centering
	\includegraphics[width=0.48\textwidth, clip, trim=6 7 8 10]{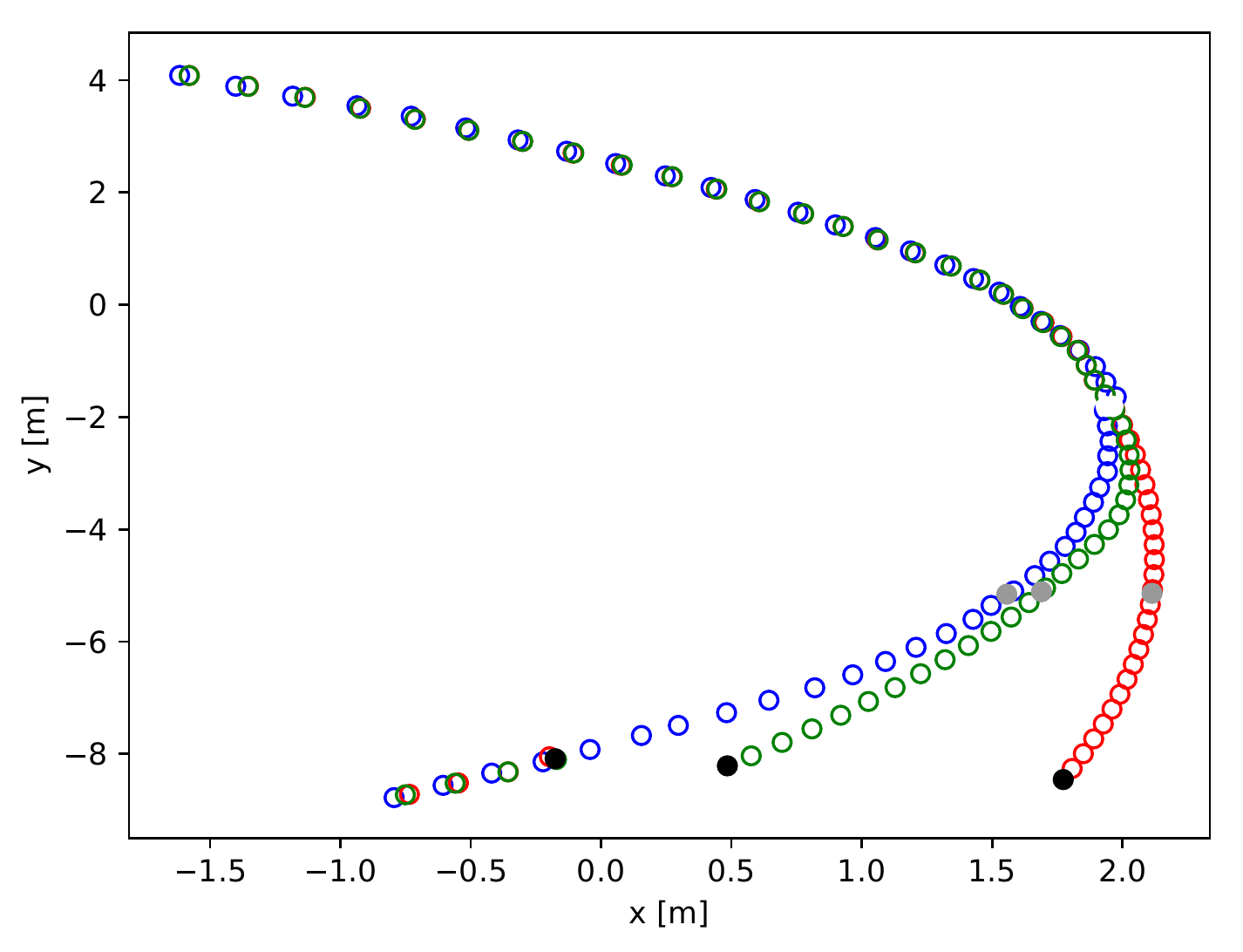}
	\vskip 2mm
	\caption{Example of a turning right scenario under a \SI{2}{\second} occlusion with model $\mathscr{C}$ (green) following the ground truth trajectory (blue) closely in contrast to model $\mathscr{P}$ (red).}
	\label{fig:goodExample}
	\vskip -0.3cm
\end{figure}

Fig.~\ref{fig:goodExample} shows an example scene for turning right with a \SI{2}{\second} occlusion. For visibility reasons only every $4^{th}$ frame in $x$- and $y$-direction is plotted. The coordinate system is the local one at the intersection and the units are given in meters. A circle represents a single position in a track. Blue represents the ground truth, green the model $\mathscr{C}$ and red the model $\mathscr{P}$ track. The filled circles of a single gray-scale tone mark the positions of the three tracks at the same time stamp to visualize velocity differences. The white filled circles mark the start of the occlusion. The green track follows the ground truth closely, but looking at the synchronization points, it slightly falls back. Considering the visualizations like in Fig.~\ref{fig:goodExample} for all turning right scenes, the velocity estimates received by the smart devices tend to have a delay when it comes to acceleration. Still, the combined model manages to track the cyclist quite accurately despite the occlusion. The intersection only model is unable to do so.

\begin{figure}[htb]
	\centering
	\includegraphics[width=0.48\textwidth, clip, trim=6 7 8 10]{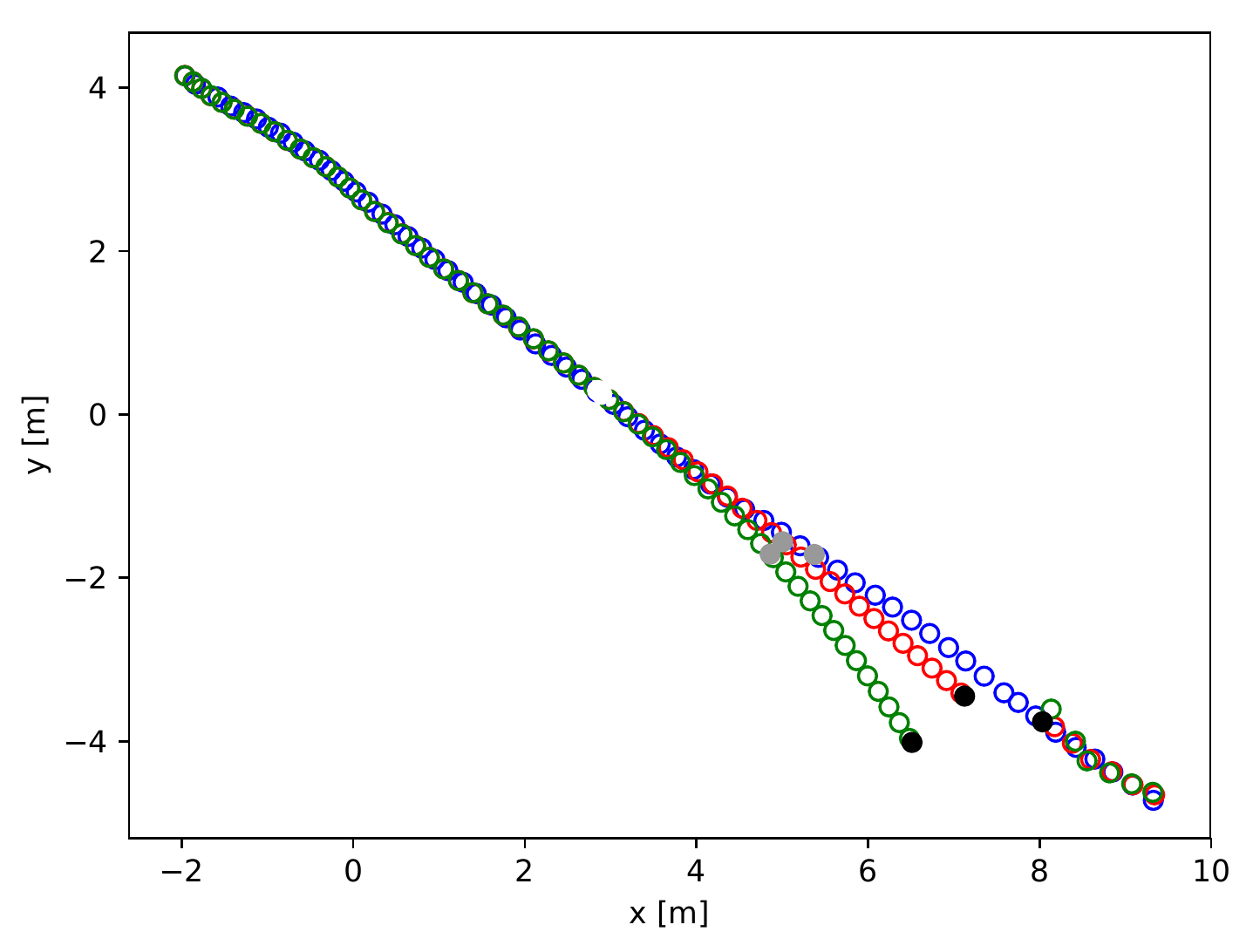}
	\vskip 2mm
	\caption{Example of a starting scenario under a \SI{2}{\second} occlusion with model $\mathscr{P}$ (red) performing better in following the ground truth track (blue) than model $\mathscr{C}$ (green).}
	\label{fig:badExample}
	\vskip -0.3cm
\end{figure}

In Fig.~\ref{fig:badExample} an example is shown with model $\mathscr{C}$ drifting apart from the ground truth track for a moving straight scene. Imprecise yaw rate data leads to the drift. As model $\mathscr{C}$ is under occlusion purely relying on smart device data, it is sensible to imprecise data. Model $\mathscr{P}$ performs better, as the direction does not change significantly under occlusion.

{
	\setlength{\abovecaptionskip}{5pt plus 3pt minus 3pt}
	\begin{table}[htb]
		\begin{center}
			\caption{MOTAP of scenes under artificial occlusions with ground truth assignment.} 
			\begin{tabular}{| c | c | c | c |}
				\hline
				Scene Type & Occlusion[s] & $\Sigma\, \text{MOTAP}(\mathscr{P}, \mathscr{C})$ & $\Sigma\, \text{MOTAP}(\mathscr{C}, \mathscr{P})$ \\ \hline
				Starting & 1 & 8 & 18 \\\hline
				Starting & 2 & 19 & 30 \\\hline
				Turning & 1 & 3 & 37 \\\hline
				Turning & 2 & 9 & 52 \\\hline
			\end{tabular}
			\label{tab:MOTAP_AssignmentEval}
		\end{center}	
		\vspace{-3mm}
	\end{table}
}

The presented results are performed under the association of the smart device data to the detected tracks. The association has to cope with multiple simultaneous cyclist tracks in 18 starting scenes and 33 turning scenes. In the other scenes, only the equipped cyclist was in the scenes and no other cyclists were falsely detected. Overall 97.7\% of the assignments in the starting scenes and 95.9\% in the turning scenes are assigned correctly. Tab.~\ref{tab:MOTAP_AssignmentEval} shows that the performance of model $\mathscr{C}$ measured with MOTAP is only slightly better under a perfect assignment.

GNSS data is used by the smart devices to estimate the velocity. The position information in combination with the horizontal dilution of precision (HDOP) can be included as additional measurement for the update of the EKF and in the assignment process. Experimental results showed no significant gain in the presented performance of the combined model. The reason is the bad quality of the GNSS data in the intersection area.
\section{\large Conclusions and Future Work}
\label{sec_conclusion}

													
In this article, we presented an approach to cooperatively track cyclists. 
The cooperation combined smart device information with an infrastructure based detection to improve the infrastructure only tracking of cyclists.
We showed by evaluation of real traffic starting and turning right scenarios using MOTA, MOTP, and the novel MOTAP measure that the addition of smart device information leads to a better tracking of cyclists in terms of accuracy and robustness. We assumed an ideal communication medium with negligible delay, but operated with real smart device sensor data.

Our future work will focus on the improvement of the accuracy of smart device data. Especially the velocity estimation could be more precise during acceleration and deceleration. This article concentrated on the use of smart device data. In a next step, we will transfer infrastructure information to smart devices to improve the self-localization methods of smart devices via intersection data. A more precise self-localization ability of the smart devices is necessary to be able to use the position for sensor fusion or in the assignment process. So far, the pure GNSS localization data is not precise enough to lead to a gain. Nevertheless, the current association algorithm showed good results. Although the test scenes were recorded in public traffic, challenging situations with multiple cyclists were rare. Our aim is to record more challenging scenes with several cyclists equipped with smart devices and to develop a more robust assignment algorithm based on a cooperative approach.

To be able to evaluate the gain of our approach with as few simplifying assumptions as possible, we will implement a realistic communication medium in further research, getting us closer towards our envisioned future traffic scenario~\cite{BRZ+17}.


\section{\large Acknowledgment}

This work results from the project DeCoInt$^2$, supported by the German Research Foundation (DFG) within the priority program SPP 1835: "Kooperativ interagierende Automobile", grant numbers DO~1186/1-1, FU~1005/1-1, and SI~674/11-1. Additionally, the work is supported by “Zentrum Digitalisierung Bayern”.



\bibliographystyle{IEEEtran}
%

{\small
\bibliography{IEEEabrv,egbib,sz,mb,gr}
}

\end{document}